\documentstyle[aas2pp4]{article}

\received{13 July 1995}
\lefthead{RYDER ET AL.}
\righthead{RINGED BARRED GALAXIES}

\slugcomment{Submitted to {\it The Astrophysical Journal}.}

\begin{document}

\title{Neutral Hydrogen in the Ringed Barred\\
       Galaxies NGC~1433 and NGC~6300\altaffilmark{1}}

\author{\sc Stuart D.~Ryder\altaffilmark{2}, Ronald J.~Buta,
       Hector~Toledo\altaffilmark{3}, and Hemant Shukla}
\affil{Department of Physics and Astronomy, University of Alabama,
       Box 870324,\\ Tuscaloosa, AL 35487-0324\\
       Electronic mail: sdr@newt.phys.unsw.edu.au; buta@sarah.astr.ua.edu}

\author{\sc Lister Staveley-Smith}
\affil{Australia Telescope National Facility, CSIRO, P. O. Box~76,
       Epping, NSW~2121, Australia\\
       Electronic mail: lstavele@atnf.csiro.au}

\and

\author{\sc Wilfred Walsh}
\affil{School of Physics, University of New South Wales, Sydney~2052,
       Australia; and Australia Telescope National Facility,
       CSIRO, P. O. Box~76, Epping, NSW~2121, Australia\\
       Electronic mail: wwalsh@atnf.csiro.au}

\altaffiltext{1}{Based on observations made with the Australia Telescope
                 Compact Array, which is operated by CSIRO Australia Telescope
                 National Facility.}
\altaffiltext{2}{Current address: School of Physics, University of New
                 South Wales, Sydney 2052, Australia.}
\altaffiltext{3}{Also at Instituto de Astronomia-UNAM, M\'{e}xico.}

\begin{abstract}
We have made observations of the \ion{H}{1} in the southern ringed barred
spiral galaxies NGC~1433 and NGC~6300 with the Australia Telescope Compact
Array (ATCA), the main goal being to test the resonance theory for the
origin of these rings. NGC~1433 is the prototypical ringed barred spiral,
and displays distinct \ion{H}{1}~counterparts to its nuclear ring, inner
ring, outer pseudoring, and plume-like features. The $L_{4}$ and $L_{5}$
regions at corotation, as well as the bar itself, are relatively devoid
of neutral atomic hydrogen. We use the Tully-Fisher relation to argue
that the mean inclination of the disk of NGC~1433 is closer to $33^{\circ}$
than to $20^{\circ}$, meaning that its outer pseudoring is intrinsically
almost circular, while the inner ring is rather more elongated than the
average (based on results from the {\em Catalog of Southern Ringed Galaxies}).
Strong radio continuum emission is localized to the nucleus and the ends
of the bar in NGC~1433, and we place an upper limit on the 1.38~GHz flux
of the Type~II SN~1985P a decade after the explosion. By associating the
inner ring of NGC~1433 with the inner second harmonic resonance, and its
outer pseudoring with the outer Lindblad resonance, we are able to infer
a bar pattern speed for NGC~1433 of $26\pm5$~km~s$^{-1}$~kpc$^{-1}$.

By way of contrast, NGC~6300 possesses a much more extended \ion{H}{1}~disk
than NGC~1433, despite having a similar morphological type. There is a
gas ring underlying the inner pseudoring, but it is both broader
and slightly larger in diameter than the optical feature. The outer
\ion{H}{1}~envelope has a $20^{\circ}$ kinematic warp as well as a
short tail, even though there are no nearby candidates for a recent
interaction with NGC~6300. The non-circular motions inferred from
optical emission-line spectra do not appear to extend beyond the bar
region of NGC~6300. Barely 10\% of the 1.38~GHz radio continuum emission
in NGC~6300 originates in the type~2 Seyfert nucleus, with the rest
coming from a disk-like component internal to the ring. By again linking
an inner ring feature to the inner second harmonic resonance, we derive
a bar pattern speed for NGC~6300 of $27\pm8$~km~s$^{-1}$~kpc$^{-1}$, but
in this case, neither the outer pseudoring nor the nuclear ring predicted
by the resonance-ring theory can be identified in NGC~6300. Although it
may be the case that the ring in NGC~6300 is not related to a resonance
with the bar at all, we postulate instead that NGC~6300 is merely a less
well-developed example of a resonance-ring galaxy than is NGC~1433.

\end{abstract}


% That's it for the front matter.  On to the main body of the paper.
%

\section{Introduction}

The barred (SB) and weakly barred (SAB) systems comprise about
two-thirds of the disk galaxy population (de~Vaucouleurs~1963).
Hydrodynamical models have demonstrated that even a mild oval distortion in
the stellar mass distribution can generate spiral structure in the gas,
leading to shocks, significant non-circular motions (up to 100~km~s$^{-1}$
in the bar region), and secular redistribution of the disk material, effects
which should be especially pronounced if the bar is strong (Sellwood \&
Wilkinson 1993, and references therein). Furthermore,
the interaction between a bar in the mass distribution and its principal
resonances in the disk can lead to distinctive concentrations of gas into
ring-like structures, whose shapes and orientations with respect to the bar
depend on the characteristics of the main periodic orbits in the resonance
regions (Schwarz 1981).

Optical observations to date of the nuclear, inner, and outer rings and
pseudorings occasionally observed in barred galaxies have strongly
suggested that these features are almost certainly manifestations of three
principal orbital resonances in the disk; the inner Lindblad resonance (ILR),
inner second harmonic resonance (2HR$^{-}$), and outer Lindblad resonance
(OLR), respectively (Athanassoula et al. 1982; Kormendy 1982; Buta 1984,
1986a, b). These resonances occur where the bar pattern speed,
$\Omega_{\rm p}$, is equal to $\Omega - \kappa / 2$, $\Omega - \kappa / 4$,
and $\Omega + \kappa / 2$, respectively (where $\Omega$ is the mean angular
velocity, and $\kappa$ is the radial epicyclic frequency). If these
identifications are correct, then the rings can, in principle, tell us the
locations of the main bar resonances, and in conjunction with the disk
kinematics, allow us to directly measure the bar pattern speed,
a fundamental unknown of galactic dynamics.

Furthermore, when the bar forcing (actually, a torque) exceeds some
critical value ($\sim20\%$ of the mean axisymmetric force), then
orbits near the Lagrangian points, $L_{4}$ and $L_{5}$ at the
corotation radius (CR, where $\Omega = \Omega_{\rm p}$) will be swept
almost completely clean (Schwarz 1981; Contopoulos \& Papayannopoulos
1980). Since periodic orbits in a bar potential tend to be aligned
parallel or perpendicular to the bar and achieve maximum local
eccentricity in the principal resonance regions (Contopoulos 1979),
then rings and pseudorings which form in these regions tend to be oval
in shape and are aligned either parallel or perpendicular to the
bar. This leads to significant non-circular motions, the amplitude of
which can be used to measure the amplitude of the bar forcing.

Such effects ought to be most easily observed in the gas distributions
of these galaxies, but so far, only a handful have been mapped in
\ion{H}{1}.  Studies of the ringed galaxies NGC~1291 and NGC~5101
(van~Driel, Rots, \& van~Woerden 1988), NGC~1365 (J\"{o}rs\"{a}ter \&
van~Moorsel 1995, hereafter JvM), NGC~1398 (Moore \& Gottesman 1995),
NGC~4725 (Wevers 1984), NGC 4736 (Mulder \& van~Driel 1993), NGC~5850
(Higdon \& Buta 1995), and NGC~2273, NGC~4826, and NGC~6217 (van~Driel
\& Buta 1991) have generally found concentrations of \ion{H}{1}~gas
coincident with the optical rings and pseudorings, together with a
pronounced central hole, at least as large as the bar (if
present). However, none of these galaxies possesses the full range of
ring types, and a patchy knowledge of their kinematics has made it
difficult to rigorously test the association between rings and
resonances.

Until quite recently, two of the very best examples of ringed barred
galaxies were inaccessible to existing aperture synthesis telescopes.
The southern spiral NGC~1433, besides being one of the nearest,
largest, and best presented examples, is also the prototype for the
class of ringed, barred spiral galaxies. It possesses all of the 3
main ring types normally found (a nuclear ring/lens, an inner ring,
and an outer pseudoring), as well as a remarkable pair of ``plumes'',
or short spiral arcs symmetrically leading the bar. In a comprehensive
study, Buta (1986b; hereafter B86) used multi-color surface photometry
and H$\alpha$ Fabry-Perot interferometry to study the star formation
and kinematic properties of these rings. On kinematical grounds, he
was able to associate each of the ring features with the orbital
resonances already mentioned. Owing to the rather localized
distribution of ionized material however, it was not possible to carry
out a complete dynamical analysis. Of particular importance in that
study was kinematic verification of the extreme intrinsically oval
shape of the inner ring, and its alignment parallel to the bar. At the
time of these observations, rings were thought to be nearly circular,
and the significance of alignments was not yet appreciated (see
Kormendy 1979, 1982).

Despite also being classed as an intermediate-type ringed barred
spiral, NGC~6300 makes for an interesting contrast with NGC~1433,
having a virtually identical angular size and distance. There is no
nuclear ring, and the outer arms do not form an outer pseudoring (Buta
1987; hereafter B87). The inner ring is exceptionally broad and
dusty, and may in fact not be aligned parallel to the bar, as is the
case for NGC~1433. The rotation curve, derived from optical emission
line spectra, shows evidence for non-circular motions in the bar,
consistent with bar-driven gas inflow. Interestingly, the rotation
curve within the optical disk is remarkably well predicted by the
$V$-band light distribution, using only a single constant value for
the mass-to-light ratio.

We have used the Australia Telescope Compact Array (ATCA) to map the
\ion{H}{1} line emission and the 1.38~GHz continuum emission in both
of these galaxies. In this paper, we describe the ATCA observations in
\S~\ref{s:atca}, present the results of our analysis in
\S~\ref{s:res}, and then discuss our findings in relation to the
resonance hypothesis in \S~\ref{s:disc}. A list of fundamental and
observational parameters for NGC~1433 and NGC~6300, along with new
quantities derived by us in this study, is given in
Table~\ref{t:obspar}.

\section{\label{s:atca}ATCA Observations}

Both galaxies were observed using a number of separate ATCA configurations
between 1994 March 2 and 1994 June 4, typically for $\sim12$~hours
each per configuration. The baselines employed spanned the range from 31~m
to 1500~m. Some details of the observations
are summarized in Table~\ref {t:obspar}. For the \ion{H}{1} observations,
a bandwidth of 8~MHz was used, divided into 512 channels each covering
3.3~km~s$^{-1}$, and centered on a frequency of 1.415~GHz (corresponding
to a heliocentric velocity in the optical convention to be used here
of $c\Delta\nu/\nu = 1146$~km~s$^{-1}$). The systemic velocities of NGC~1433
and NGC~6300 as measured by Reif et al. (1982) with the Parkes
telescope (1080 and 1110~km~s$^{-1}$, respectively) are close enough that
we were able to use the same observing parameters for both. The second IF
was used to simultaneously record measurements of the continuum emission
within a 32 channel/128~MHz bandwidth centered on 1.38~GHz.

The primary ATCA calibrator, PKS B1934--638, was normally observed at
least once per 12~hour run, assuming a flux density of 16.2~Jy at
1.4~GHz.\footnote{The recent recalibration of the 1934--638 flux
density scale by Reynolds (1994) would revise this value
(and all subsequent fluxes and \protect{\ion{H}{1}} mass
determinations) downwards by 8\%.} Each 40 minute scan of a galaxy was
bracketed by short observations of a nearby secondary calibrator:
PKS~B0438--436 ($\sim5.2$~Jy at 1.4~GHz) for NGC~1433, and
PKS~B1718--649 ($\sim3.8$~Jy at 1.4 GHz) for NGC~6300.  Each
$uv$-database for a given configuration, galaxy, and IF (spectral line
or continuum) was processed separately using implementations of the
NRAO {\sc aips} package at the Australia Telescope National Facility
and at the University of Alabama. After interactive removal of
spurious data from both polarizations on each of the 10 baselines, the
data were corrected for gain and phase variations, and then a bandpass
calibration based upon observations of the primary calibrator was
applied. The spectral line data were then further processed to remove
any underlying continuum, the original 20~second integrations were
extended to 60~seconds by averaging, and finally, Doppler corrections
were applied to bring the individual configurations to a common
velocity reference frame.

Data from all of the available configurations were combined together
into a single $uv$-plane (in the case of the continuum data) using
bandwidth synthesis, or (in the case of the spectral line data)
channels were averaged together in groups of 3, separated by
9.9~km~s$^{-1}$. Each continuum map (or plane of a spectral line
cube), consisting of $256\times256\ 10''$ square pixels, was CLEANed
until 1000 iterations were completed, or until the total cleaned flux
density changed by less than 0.1\% (usually the former). Two separate
weighting schemes were employed: uniform weighting, yielding a
synthesized beamwidth of $\sim 30''$, and natural weighting, giving a
beam nearly twice as large, but with an expected improvement in
surface brightness sensitivity. Figure~\ref{f:beamplot} compares the
response of the ``dirty'' natural-weighted beam with that of the
restoring ``clean'' beam (in essence, a Gaussian fitted to the central
peak of the dirty beam), as well as the integrated flux measured by
each type of beam (see the appendix of JvM for a discussion of the
effects of the beam shape on cleaning and integrated fluxes). After
correcting the data cubes for the effects of primary beam attenuation,
moment maps (\ion{H}{1} column density, velocity, and velocity width)
were constructed using a mask derived from a spatially and spectrally
smoothed version of each cube, and by setting a flux cutoff that just
enabled a clean separation of the \ion{H}{1} emission from background
noise.

\section{\label{s:res}Results}

\subsection{\label{s:n1433}NGC~1433}

\subsubsection{\label{s:n1433h1d}H\,I gas distribution}

Figure~\ref{f:n1433cm} shows each of the channel maps in the
uniform-weighted cube, identified by their respective heliocentric
velocities. To first order, NGC~1433 shows the characteristic pattern
of uniform circular rotation, although there are clearly gaps in the
\ion{H}{1} distribution. To determine the extent to which we may be
missing \ion{H}{1} flux due to missing $uv$-spacings (or perhaps
overestimating the flux due to inadequate cleaning; JvM), we have
measured global \ion{H}{1} profiles from both the uniform- and
natural-weighted cubes, using a moving window to optimize the emission
measured in each individual channel map. The two profiles are shown
together in Figure~\ref{f:n1433prof}, together with a new spectrum
obtained recently by us with the Parkes 64~m
telescope. Table~\ref{t:profpars} compares our measurements of the
moments for each of these three global \ion{H}{1} profiles.

Despite using a total of 4~array configurations to increase our
sampling in the $uv$-plane, we would still expect to resolve out some
of the lowest surface brightness emission. Ryder et al. (1995), using
a total of 5~ATCA configurations, found a 6\% loss of total flux for
the late-type barred spiral NGC~1313, compared with the integral from
a grid of Parkes observations. The flux integral measured from the
uniform-weighted cube is 32.2~Jy~km~s$^{-1}$, while that from the
natural-weighted data is only slightly more at 34.4~Jy~km~s$^{-1}$.
Both of these flux integrals agree well with our new Parkes
measurement, which represents nearly a 50\% increase in total
\ion{H}{1} content over that measured by Reif et al. (1982) from a
relatively poor signal-to-noise spectrum of NGC~1433. Our Parkes
spectrum has not been corrected for dilution by the primary beam, but
the extent of the \ion{H}{1} disk does not appear to be sufficient to
cause overfilling of either the Parkes beam ($15'$ at 1.4~GHz), or the ATCA
primary beam ($34'$).

Adopting a distance to NGC~1433 of 11.6~Mpc (Tully 1988, but see
\S~\ref{s:inc}) and assuming optically thin gas, our measurements of
the flux integral in NGC~1433 yield a total \ion{H}{1} mass of
$(1.05\pm0.05)\times10^{9}$~M$_{\odot}$. For an absolute blue
magnitude of $M_{B}^{b,i=33}=-19.21\pm0.33$ (\S~\ref{s:inc}), then the
ratio $M_{\rm H\,I}/L_{B}=0.14\pm0.05$~M$_{\odot}/L_{\odot}$. This is
well within the median range of \ion{H}{1}~mass-to-blue luminosity
ratios found in Sab galaxies by Roberts \& Haynes (1994), so that even
though the distribution of gas and stars in NGC~1433 is somewhat
unusual, the relative \ion{H}{1} content is not.

Since we do not appear to be losing any significant emission by using
uniform weighting, we shall discuss only the uniformly-weighted data
from here on.  The global profile width at 20\% of the peak level
($W_{20}$), uncorrected for either turbulent or instrumental
broadening effects, is $190\pm10$~km~s$ ^{-1}$
(Table~\ref{t:profpars}). Unlike the \ion{H}{1} flux integral, this
value is in extremely good agreement with that measured by Reif et
al. (1982). We shall defer the conversion of $W_{20}$ to a deprojected
velocity width until \S~\ref{s:inc}, after we have had a
chance to consider the \ion{H}{1} morphology and kinematics.

The \ion{H}{1} column density distribution is shown as both a
grey-scale and as contours in Figure~\ref{f:n1433h1}. The Full-Width
at Half-Power (FWHP) of the elliptical Gaussian restoring beam is
shown in the lower-left of this figure. The tight confinement of the
neutral atomic hydrogen to both an inner and an outer ring is quite
striking. Just how good the alignment really is between the gas and
the stellar rings is best illustrated in Figure~\ref{f:h1onb}, where
the \ion{H}{1} column density contours have been overlaid on a 600~s
$B$-band CCD image of NGC~1433 obtained with the CTIO 1.5~m telescope.
The $B$~image has been placed on a logarithmic scale in order to bring
out the faint outer pseudoring as well as the inner ring, bar, and
nuclear ring. Comparison of Fig.~\ref{f:h1onb} with
Fig.~\ref{f:n1433h1} shows that both the inner and the outer stellar
rings are almost perfectly aligned with their atomic gas
counterparts. The gas density peaks near the ends of the bar, where it
joins the major axis of the aligned inner ring. Neutral gas is
deficient in the bar itself, but appears to build up again at a point
just slightly to the east of the optical nucleus (which is itself
virtually coincident with the dynamical center as determined in
\S~\ref{s:n1433h1k}). Secondary peaks of \ion{H}{1} surface density
are seen to underlie both of the ``plume'' features discussed by B86.
Areas in between the bar and the inner ring, and especially regions
between the inner ring minor axis and the outer pseudoring, are
relatively devoid of neutral hydrogen.

The azimuthally-averaged radial variation of the \ion{H}{1} column density
has been calculated using the IRING task, by assuming an inclination of
$33^{\circ}$ (\S~\ref{s:inc}) and other disk parameters taken from the
kinematical analysis (\S~\ref{s:n1433h1k}). The radial profile of \ion{H}{1},
corrected to face-on, is shown in Figure~\ref{f:n1433ir}. Also marked in
this figure by dashed lines are the mean apparent radii for the various
morphological features, as tabulated by B86. The \ion{H}{1}
surface density starts from a peak of 2~M$_{\odot}$~pc$^{-2}$ in the nucleus,
and drops to as low as 1~M$_{\odot}$~pc$^{-2}$ along the bar, before reaching
its highest value of some 2.5~M$_{\odot}$~pc$^{-2}$ just outside the mean
radius of the inner ring. Despite localized \ion{H}{1} peaks underlying the
plumes, the mean gas surface density at their radial distance is lower
overall because of the ``missing'' gas in the inter-ring region. The
\ion{H}{1} column density reaches one more peak at or just beyond the radius
of the outer pseudoring, before declining quasi-exponentially thereafter.
We perhaps ought not to be too surprised at finding peaks in the gas surface
density just outside the stellar rings, as neither can exist precisely at
the resonant orbit radii. In the rush to escape the unstable resonant orbits,
the gas clouds collide, form new stars, and lead to an eventual buildup in
stellar surface density near to, {\em but not right at}, the resonance radii.
We note in passing that the outermost point at which the \ion{H}{1} surface
density drops to a level of 1~M$_{\odot}$~pc$^{-2}$ corresponds to a distance
of 1.18$R_{25}$ ($R_{25}=D_{25}/2$), quite typical for Sa--Sab galaxies
(Warmels 1986) despite the perturbing actions of the bar and its
associated resonances.

\subsubsection{\label{s:n1433h1k}H\,I Kinematics}

The \ion{H}{1} velocity field, as determined from the moment analysis,
is shown superimposed on the gas column density map in Figure~\ref{f:n1433vf}.
Owing to the low gas densities in the outskirts of the disk, and between
the inner and outer rings, we have been forced to mask out many unreliable
data points in these areas. Even so, we can still get a good feel for how
complex the velocity field in this particular galaxy really is. The ``spider''
pattern characteristic of disks in uniform circular rotation is still evident,
but modulated heavily by the non-uniform gas distribution, and apparent
departures from circular motion. There is an exceptionally steep velocity
gradient along the central $1'$ or so of the kinematic major axis (roughly
perpendicular to the bar), with severe distortions along the southern edge
of the optical bar. Further out along the major axis, the isovelocity contours
close up when they reach the inner ring, implying that the rotational velocity
reaches a maximum there, and then begins to drop. Somewhat perversely, a lack
of gas between the inner and outer rings with which to trace the motions makes
the velocity field somewhat chaotic in this important transition region.
Fortunately, there is plenty of gas along the kinematic minor axis, although
there the correction for deprojection will introduce a large uncertainty. In
regions of high gas surface density, there are kinks indicative of
non-circular motions, but with no systematic pattern to them.

We have attempted to determine a rotation curve from the velocity field
using an implementation of the ROCUR algorithm (Begeman 1987)
within {\sc aips} to fit elliptical rings of varying inclination
and position angle. As starting values for the kinematical parameters
(dynamical center position, systemic velocity, and disk tilt and position
angle), we adopted the numbers arrived at by B86 in his
photometric and optical interferometric study of NGC~1433. However
the gas deficiencies along the major axis, as well as the need to exclude
velocity points within $\pm25^{\circ}$ of the minor axis, tend to result
in convergence towards unreasonably small values for the disk inclination
(and correspondingly excessively large deprojected rotation velocities) if
all the parameters are allowed to vary freely. The only way to get a
satisfactory fit to the velocity field is to assume {\em a priori\/} a constant
inclination for the gas orbits at all radii.

By assuming the outer isophotes are intrinsically circular, B86
arrived at an inclination of $i=33^{\circ}$. On the other hand, statistics
from the Catalog of Southern Ringed Galaxies (Buta 1995, hereafter
CSRG) indicate that outer ring features in strongly barred galaxies have,
{\em on average}, an intrinsic axial ratio of $0.82\pm0.07$. For this reason,
and others to be discussed in \S~\ref{s:inc}, he conceded that
the actual inclination of NGC~1433 to the plane of the sky could be as low as
$20^{\circ}$ or so. Figure~\ref{f:n1433h1} is not immediately helpful in
regard to settling this matter, owing to the way the gas is so clearly
organized into these same rings. The lowest contour in Fig.~\ref{f:n1433h1}
(1.1~M$_{\odot}$~pc$^{-2}$) spans roughly $7\farcm5\times6\farcm3$ ($25
\times21$~kpc at a distance of 11.6~Mpc), yielding an axial ratio $q=0.84$,
wholly consistent with the isophote axis ratio $q=0.85$ measured by B86.

Regardless of whether the inclination is held fixed at either $33^{\circ}$ or
at $20^{\circ}$, the ROCUR analysis does produce consistent results for
all the other parameters, if they are allowed to vary freely. We find a
systemic velocity $V_{\rm hel}=1075\pm2$~km~s$^{-1}$, and a kinematic major
axis position angle of $\theta=201^{\circ}\pm1^{\circ}$. These results
compare quite favorably with the equivalent parameters determined by
B86 from his optical kinematic modeling: $V_{\rm hel}=1079\pm2$~km~s$^{-1}$
and $\theta=197^{\circ}$, and with the systemic velocities measured from the
global \ion{H}{1}~profiles (Table~\ref{t:profpars}). The dynamical center
of the \ion{H}{1} disk lies at $\alpha_{\rm dyn}({\rm J2000}) = 03^{\rm h}\,
42^{\rm m}\, 1\fs5\, (\pm10'')$, $\delta_{\rm dyn}({\rm J2000}) = -47^{\circ}
\, 13'\, 12''\, (\pm5'')$. Within the error bars, this is virtually coincident
with the position of the optical nucleus as given by
de~Vaucouleurs et al. (1991; hereafter RC3). As mentioned
in the previous section, the central peak in the \ion{H}{1}
distribution is offset to the east of the dynamical center, though by
less than one synthesized beamwidth.

In \S~\ref{s:inc}, we demonstrate that the most plausible inclination for the
disk of NGC~1433 as a whole is of order $33^{\circ}$. By holding all other
parameters fixed as above, and finding the mean rotational velocity within
annuli of increasing semi-major axis, we have compiled separate rotation
curves for the approaching and receding halves, as well as for the disk
as a whole. Figure~\ref{f:n1433rot} shows plots of these deprojected
rotation velocities. Any differences between the two halves are a better
reflection of velocity uncertainties than the formal fitting errors.
Clear asymmetries in the velocity field around the bar visible in
Fig~\ref{f:n1433vf} lead to a large spread in possible values within
the first $1'$ or so, but further out, agreement between the approaching
and receding halves rapidly improves. After an extremely steep rise to a
maximum rotational velocity of about 170~km~s$^{-1}$ just inside the
bar/inner ring region, the rotation curve levels off at just on 150~km~s$
^{-1}$, even as it passes the radius of the outer pseudoring ($174''$;
B86). A model velocity field was constructed from the
full-disk rotation curve, and subtracted from the observed velocity field.
The r.m.s.~residuals are only of order 10~km~s$^{-1}$, but exhibit roughly
a $\sin3\phi$ variation with azimuthal angle $\phi$, suggesting that a
fixed inclination of $33^{\circ}$ may still not be appropriate everywhere in
the disk.

If anything, the central velocity gradient is likely to be even
steeper than shown in Fig.~\ref{f:n1433rot}, due to the smoothing
effect of our $\sim30''$~beam. H$\alpha$ Fabry-Perot kinematics would
be an ideal complement to the \ion{H}{1}~velocity field, adding
spatial resolution where it is needed most. Unfortunately, the
predilection of the \ion{H}{2}~regions for the inner and nuclear rings
alone made it impossible for B86 to trace the velocity field in
between these rings. With the increased sensitivity of modern
instrumentation to lower surface brightness emission however (e.g.,
Bland-Hawthorn et~al.  1994), it may yet be possible to derive
kinematical information from this critical region.

\subsubsection{\label{s:n1433c}Radio continuum}

As part of a general imaging program of southern ringed galaxies,
Crocker, Baugus, \& Buta (1995) have used the CTIO 1.5~m
telescope to obtain images in the narrow-band H$\alpha$ and adjacent
red continuum. Figure~\ref{f:n1433conha} shows a 1200~s continuum-subtracted
H$\alpha$ image of NGC~1433, on which has been overlaid contours of
the 1.38~GHz continuum emission (uniform-weighted) observed with the
ATCA. Not surprisingly, most of the radio continuum contours outline
quite well the star-forming complexes found at each end of the bar, as
well as the plumes and the nuclear ring. The integrated flux densities
from the \ion{H}{2}~region complexes at the eastern and western ends
of the bar are found to be 3.4~mJy and 3.7~mJy respectively,
comparable to the total nuclear flux density of 3.4~mJy. In the
absence of observations at a second frequency, we can only speculate
on the source of this emission, but we expect that any non-thermal
component will dominate the continuum emission at 1.38~GHz
(Condon 1992). Thus, the concentrated emission originating
near the ends of the bar and the inner ring minor axis is presumably
associated more with supernovae (and/or focusing of the magnetic field
lines due to the orbit crossings there) than with the enhanced star
formation activity that eventually gives rise to the supernova
activity. Radio continuum emission is much sparser in the outer
pseudoring, as indeed is the number of \ion{H}{2}~regions. The origin
of the westernmost arc of continuum emission is unclear, as it lies
outside any optical or \ion{H}{1}~feature.

A growing number of Type~II supernovae have now been detected as radio
and X-ray sources within 10--20 years of the optical outburst, e.g.,
SN~1986J (Rupen et~al. 1987) and SN~1978K (Ryder et~al. 1993). The cross in
Fig.~\ref{f:n1433conha} to the northeast of the nucleus of NGC~1433
marks the best known optical position of the Type~II supernova SN~1985P
(Barbon et~al. 1989). Within the $30''$ beamsize, this
cross coincides with the peak of an extended patch of emission along
the inner ring of NGC~1433, and at least one weak H$\alpha$ source.
Although this identification is far from definite, we can at least
place an upper limit on the integrated flux density of SN~1985P at the
current epoch as $S_{1.38~{\rm GHz}} \lesssim 0.5$~mJy. This places
SN~1985P at least an order of magnitude fainter than the ``normal''
Type~II radio supernova SN~1979C at the equivalent point in its radio
evolution, and well short of the luminosities attained by the
``peculiar'' radio supernovae exemplified by SN~1986J and SN~1978K
(Weiler et~al. 1995). Based upon our present understanding of the
evolution of young and intermediate-age supernova remnants (e.g.,
Chevalier \& Fransson 1994), such a low radio luminosity nearly
a decade after outburst would require a relatively low progenitor
mass-loss rate, and a correspondingly rarefied circumstellar medium
for the expanding supernova to interact with.

\subsection{\label{s:n6300}NGC~6300}

\subsubsection{\label{s:n6300h1d}H\,I gas distribution}

As was the case with NGC~1433, contour plots of the uniform-weighted
channel maps of NGC~6300 (Figure~\ref{f:n6300cm}) are to first order,
consistent with uniform circular motion, but hint at a non-uniform
surface density distribution, and some kinematic irregularities in
the outer disk. Before examining the moment maps formed from these
channel maps, we ought to again compare the global \ion{H}{1}~profiles
from the uniform- and natural-weighted datasets with an equivalent
Parkes spectrum, kindly supplied to us by M.~Mazzolini. All three
profiles are shown together in Figure~\ref{f:n6300prof}, and display
remarkably good agreement in profile width, alignment, and structure.
Table~\ref{t:profpars} shows just how well our synthesis measurements
compare with single-dish observations.

Once again, our uniform- and natural-weighted cubes yield integrated
\ion{H}{1}~fluxes that agree to better than 10\%, and come in only
slightly below the two Parkes fluxes. The new Parkes spectrum has not
been corrected for dilution due to overfilling of the primary beam,
which may be a much more significant effect than in NGC~1433, given
the larger angular extent of the gas disk in NGC~6300
(Figure~\ref{f:n6300h1}). Although we could still be missing up to 15\%
of the total flux from NGC~6300, we note that all three spectra in
Fig.~\ref{f:n6300prof} outline the double-horned profile of NGC~6300
much better than does the spectrum published by Reif et al.
(1982). Since most of the low surface brightness gas is expected to
lie in the flat regime of the rotation curve that gives rise to these
strong horns, it would not appear as though we are resolving out
significant quantities of this gas. Adopting as the distance to
NGC~6300 a figure of 14.3~Mpc (Tully 1988), we therefore estimate an
\ion{H}{1}~gas content for NGC~6300 of $(2.6\pm0.2) \times
10^{9}$~M$_{\odot}$. Thus, NGC~6300 has at least twice as much neutral
hydrogen as NGC~1433, and yet $M_{\rm H\,I}/L_{B}=0.18\pm
0.06$~M$_{\odot}/L_{\odot}$, almost identical to the ratio found in
NGC~1433 (\S~\ref{s:n1433h1d}).

Although the uniform-weighted dataset does not appear to lose significant
amounts of \ion{H}{1}~flux, our moment analysis has revealed that the
resultant map of \ion{H}{1} surface brightness fails to include some
important low surface brightness features. We show in Figure~\ref{f:n6300h1}
the grey-scale and contours of the projected gas column density in
NGC~6300, using a natural weighting of the {\em uv}-dataset that
results in a broader synthesized beam ($\sim50''$ as shown in the
lower-left corner) but an enhanced sensitivity to faint, extended
structure. Just as we found in NGC~1433, there are clear peaks in the
gas density, apparently associated with the optical stellar ring.
However, in contrast to NGC~1433, there do not
appear to be significant regions of the gas disk which have been swept
clean of gas, even allowing for the increased projection effects due
to the higher inclination of NGC~6300. Furthermore, a natural-weighted
map of the \ion{H}{1}~column density in NGC~1433 did not show such a
ragged outer disk as NGC~6300, nor did it show any feature resembling
the loose ``tail'' of gas seen emanating from the western edge of the
gas disk of NGC~6300. Although this feature is barely separable from
the background noise, a check of the associated velocity field
confirms it to be a natural extension of the main gas disk. A search
of the NASA Extragalactic Database turned up no interacting companion
candidates within a $30'$ radius of NGC~6300, suggesting that this feature is
associated with ongoing gas accretion rather than the remnant of a
past close interaction. The same velocity field also confirms the
visual impression from Fig.~\ref{f:n6300h1} of a slight clockwise
twist in the disk major axis in going from the ring to the disk edge.
The case for a warp in the disk of NGC~6300 is examined more closely in
\S~\ref{s:n6300h1k}.

For the purposes of comparing the \ion{H}{1} column density
distribution in NGC~6300 with its kinematics and stellar distribution,
we shall return to maps made from uniform-weighted data, which must
sacrifice some low surface brightness sensitivity for the fullest
possible resolution. Figure~\ref{f:n6300h1onb} shows contours of this
high-resolution gas column density map superimposed on a 600~s
$B$-band CCD image of NGC~6300. This image was also obtained with the
CTIO 1.5~m, and is displayed here on a logarithmic scale, so as to
bring out the outer optical disk, the inner ring, and the bar (the
latter partly obscured by dust lanes and by two bright field stars).
Here again, a stellar ring has been shown to have an \ion{H}{1}
counterpart, although the gas peaks in this ring appear to sit just
slightly {\em outside} the ridge line of the optical ring. Notice also that
a number of short spiral arcs, that are just discernible in the
optical image extending from the outside of the ring, also have
underlying \ion{H}{1} counterparts, in particular, one that emerges from the
eastern end of the disk major axis and wraps its way towards the west.

The gas disk as shown in Fig.~\ref{f:n6300h1} spans approximately
$14'\times8'$ (a 60~kpc diameter at 14.3~Mpc), not including the tail.
The run of deprojected \ion{H}{1}~column density with radius for the
uniform-weighted map is shown in Figure~\ref{f:n6300ir} in which the
ellipse geometries have been varied with radius according to the
best-fit rotational model of the velocity field (\S~\ref{s:n6300h1k}).
Although NGC~6300 has the same central density of \ion{H}{1} as
NGC~1433, it soon rises to over 4~M$_{\odot}$~pc$^{-2}$ in the ring,
nearly twice the maximum gas density seen in NGC~1433. Although
neither the peak of the gas ring, nor the full extent of the optical
ring are particularly well-defined, the \ion{H}{1}~ring is clearly
much broader than the stellar ring, and has a mean diameter closer to
$70''$ as compared to the optical ring diameter of $59''$ (B87;
CSRG). The small peak at $145''$ most likely reflects the southern
\ion{H}{1}~spur mentioned in the previous paragraph, although this
spike is smaller than our beamwidth, and the ring inclination is
changing rapidly in this region. From there, the \ion{H}{1} column
density slowly ramps down to our flux cutoff just past the \ion{H}{1}
tail, with no other candidates for a secondary peak that could
indicate a subtle outer ring or pseudoring. Finally, we observe that
the \ion{H}{1}~surface density falls below 1~M$_{\odot}$~pc$^{-2}$ at
a distance of $1.76R_{25}$, significantly larger than in NGC~1433 or
for Sb galaxies in general.

\subsubsection{\label{s:n6300h1k}H\,I Kinematics}

Figure~\ref{f:n6300vf} shows isovelocity contours of the velocity
field, overlaid upon a grey-scale image of the \ion{H}{1}~column
density, both of them constructed from the natural-weighted data.
The kinematics of the gas tail are consistent with an
extrapolation of the 1020 to 1080~km~s$^{-1}$ contours from the main
disk. Our early suspicions about a possible kinematic warp in the
outer gas disk of NGC~6300 are also confirmed by this figure. Closed
contours at 980 and 1240~km~s$^{-1}$ indicate that the rotational
velocity reaches a maximum somewhere outside the radius of the ring.

At the resolution of this natural-weighted data ($\sim50''$), there is
little indication of the non-circular motions detected by B87 either
in the bar, or elsewhere in the disk.  Our best possible resolution
comes from the uniform-weighted $uv$-dataset, and
Figure~\ref{f:n6300bar} shows a close-up of the inner $6'\times4'$ of
NGC~6300, with \ion{H}{1}~column densities and isovelocity contours
from this dataset. Although the bending of the isovelocity contours in
the inner $1'$ is consistent with the pattern of the optical
isovelocity contours in Fig.~12 of B87, these kinks do not extend much
beyond the bar region, or across the ring itself. Thus, at the limited
resolution of our \ion{H}{1} data, the bar-driven gas inflow usually
invoked to account for these non-circular motions appears to be
confined within the bar region itself, although it is worth noting
that the \ion{H}{1}~ring would provide a convenient reservoir of gas
to sustain this process.

Because of the high projected gas column densities over the disk of
NGC~6300 {\em cf.} NGC~1433, it was possible to carry out a full
kinematical analysis of the uniform-weighted velocity field with ROCUR
that allowed both the inclination $i$ and the position angle $\theta$
to be fitted as free parameters. After initially letting all
parameters vary freely, we found the systemic velocity to remain
steady at $V_{\rm hel} =1109\pm4$~km~s$^{-1}$, and a consistent
dynamical center located at $\alpha_{\rm dyn}({\rm J2000}) = 17^{\rm
h}\, 16^{\rm m}\, 59\fs5\, (\pm4'')$, $\delta_{\rm dyn}({\rm J2000}) =
-62^{\circ} \, 49'\, 13''\, (\pm4'')$.  Our redshift is in excellent
agreement with other \ion{H}{1}~measurements (Table~\ref{t:profpars})
and with the optical velocity of $1107\pm5$~km~s$^{-1}$ (B87).
Our dynamical center is also identical with the nuclear position listed
in RC3, although our \ion{H}{1}~data still lack the spatial resolution
necessary to confirm the $5''$ displacement noted by B87.

Having locked the systemic velocity and the rotation center at these
values, the kinematical analysis was repeated to determine the optimum
position angle, inclination, and resultant deprojected rotational
velocity at $25''$ (1~beamwidth) intervals, and the results are
plotted in Figure~\ref{f:n6300rot}. As with NGC~1433, fits were done
on the disk as a whole, as well as for the separate halves, in order
to quantify the extent of any disk asymmetry. The residuals of the
fitted velocity field from that actually observed have a tendency to
be mainly positive in the gas ring, and negative in the extended
gas disk, but in both of these regions, the r.m.s.~amplitude of these
residuals seldom exceeds 10~km~s$^{-1}$.

We also have here a rare opportunity to compare directly a rotation
curve for the \ion{H}{1} with that of the ionized gas (H$\alpha$ and
[\ion{N}{2}]), taken from Table~11 of B87. Some disagreement between
the optical and \ion{H}{1}~rotation curves is to be expected in the
inner $40''$ or so, due to the smearing of the steep central velocity
gradient by our $25''$ beam, and indeed, the first few
\ion{H}{1}~points do fall slightly below the curve of B87. The sudden
rise in the optical rotation curve, starting at $r=60''$ and
coinciding with a hump in the light profile due to the inner ring, is
however not echoed by the \ion{H}{1}~curve until $r>80''$. Even then,
much of this rise can be attributed to the jump to a lower inclination
(and a corresponding increase in the deprojection factor), while the
optical rotation curve employs a constant inclination ($i=52^{\circ}$)
and line-of-nodes position angle ($\theta=108^{\circ}$)
throughout. The optical and \ion{H}{1}~rotation curves are in much
better agreement again by the time the optical data runs out at
$r\sim130''$. The cause of the disagreement of optical and \ion{H}{1}
rotation curves near $r=75''$ is still not clear, but comparisons with
the individual optical rotation profiles of NGC~6300 along various
position angles (see Table~5 of B87) suggest that non-circular motions
associated with the massive arms constituting the inner ring are
partly causing the ``bump'' in the optical rotation curve. It is also
important to note that the optical rotation curve is based principally
on spectra in four position angles, and that one position angle,
132$^{\circ}$, had three separate spectra that carried much of the
weight in the solution. As with NGC~1433, a full survey of the ionized
gas velocity field with a Fabry-Perot interferometer may be warranted
for a better comparison.

Despite some disagreement between the receding and approaching halves
of the velocity field on the actual disk inclination at each radius,
there is remarkably good agreement on the kinematic major axis.  Once
the edge of the optical disk has been reached, the line-of-nodes
slowly but steadily twists some $20^{\circ}$ clockwise, from the
photometric major axis position angle of $108^{\circ}$ to a final
position lying due east on the sky. Such warps do not appear to be a
common feature in early-to-intermediate type galaxies (Bosma 1991);
nevertheless, NGC~6300 still conforms to the first of Briggs' (1990)
empirical ``rules'' of behavior for galactic warps, in that the onset
of the warp occurs between $R_{25}$ and $R_{26.5}$. At the same time,
the disk inclination steadily increases from $46^{\circ}$ to
$55^{\circ}$ before dropping back below $50^{\circ}$ in the outermost
regions (note that there is insufficient gas in the uniform-weighted
map at the radial distance of the \ion{H}{1}~tail to extend the
kinematic analysis out that far).  The rotation curve of the extended
\ion{H}{1}~disk falls slowly from its peak of 190~km~s$^{-1}$ to just
below 170~km~s$^{-1}$, before picking up again beyond $r=300''$ as $i$
is decreased. Although the $V$-band light distribution of B87 did a
surprisingly good job of reproducing the optical rotation curve with
just a single value for the mass-to-light ratio, there is clearly
still a need for a dark-matter halo component to sustain the high
\ion{H}{1}~rotational velocities well outside the optical disk.

\subsubsection{\label{s:n6300c}Radio continuum}

Figure~\ref{f:n6300conha} shows a 600~s continuum-subtracted H$\alpha$
image of NGC~6300 obtained in the same way as described in
\S~\ref{s:n1433c}, on which has been overlaid contours of the
(uniform-weighted) 1.38~GHz continuum emission.  As expected for a
Type~2 Seyfert, there is significant emission from the unresolved
nuclear source (coincident with the optical and the dynamical center),
amounting to some 15~mJy. Surprisingly, the integrated emission from
the galaxy as a whole totals approximately 110~mJy, meaning that the
disk component contributes fully 90\% of the total flux at 1.38~GHz.
This is unusually large for an SBb type galaxy, but is not unlike the
ratio observed in the ringed SAab galaxy NGC~4736 (Mulder \&
van Driel 1993). This marks another major difference between NGC~6300
and NGC~1433, which shows a more typical ratio of nuclear-to-total
flux density for Sa and Sab galaxies of 0.32.

Within the resolution (indicated by several point sources in the same
field as Fig.~\ref{f:n6300conha}), the morphology of the extended
continuum emission matches quite well that of the star-forming ring,
with peaks on the ring major axis, nearly perpendicular to the bar
axis. The two extensions on the southern edge have counterparts in the
natural-weighted \ion{H}{1}~map (Fig.~\ref{f:n6300h1}), only much
further out, and indeed, closer examination suggests that these
features seen in both maps are in fact low-level artifacts due to the
residual effects of a confusing source just outside the primary
beam. It may yet be possible to use the four background sources shown
in Fig.~\ref{f:n6300conha}, and some even brighter sources further
out, to self-calibrate the $uv$-dataset, and/or for
\ion{H}{1}~absorption measurements, from which a gas spin temperature
in the outskirts of the \ion{H}{1}~disk could be determined.

The extended nature of the continuum emission in NGC~6300 requires that
we consider the possibility that the \ion{H}{1}~ring seen in NGC~6300 is
purely an absorption effect, rather than an actual deficiency of \ion{H}{1}.
A comparison of Figures~\ref{f:n6300h1onb} and~\ref{f:n6300conha} (both
of which are at the same image scale) shows that the \ion{H}{1}~ring and
the extended continuum disk are co-extensive, but there is no obvious
correlation (or anti-correlation) of \ion{H}{1}~column density with
continuum flux density. The central source is only marginally resolved,
if at all, by our $25''\times23''$ beam, and yet the central \ion{H}{1}
hole spans $\sim2'$, and has an aspect ratio remarkably similar to that
of the gas disk as a whole. Thus, while the magnitude of the central
\ion{H}{1}~depression may be exaggerated slightly by the presence of
the nuclear continuum source, we are satisfied that the inner \ion{H}{1}
gas in NGC~6300 does in fact form a nearly closed and continuous ring.

\section{\label{s:disc}Discussion}

\subsection{\label{s:inc}The Inclination of NGC~1433}

In his study, B86 was unable to assign a definitive
inclination for the disk of NGC~1433, since the outermost optical
isophotes (usually assumed to be circular in the plane of the galaxy)
are influenced by the presence of the outer pseudoring,
while statistical studies for a virtually complete sample of southern
ringed galaxies (CSRG) indicate that the average outer
pseudoring in SB galaxies (though not necessarily {\em this}
particular pseudoring) has an intrinsic axial ratio $q=0.82\pm0.07$.
In the former case (ring intrinsically circular), an inclination of
$33^{\circ}$ follows, but in the latter case (ring axial ratio of
0.85, and assuming the ring major axis to be oriented perpendicular to
the bar), the inclination will be closer to $20^{\circ}$. In this
section, we aim to demonstrate that, statistics aside, the higher
inclination is the only plausible one.

We start by establishing a distance to NGC~1433. Tully (1988) proposes
a distance of 11.6~Mpc, based on a redshift consistent with ours, a
Hubble constant of 75~km~s$^{-1}$~Mpc$^{-1}$, and a Virgocentric
infall model. This distance receives some support from a variety of
avenues. NGC~1433 is included as a member of the extensive Southern
Group No.~13 (Maia, da~Costa, \& Latham 1989), many of whose other
members are also believed to have distances on the order of
12~Mpc. Semi-independent confirmation also comes from the
\ion{H}{2}~region Luminosity Function (Crocker, Baugus, \& Buta 1995)
which, for an assumed distance of 11.6~Mpc, indicates that the
brightest \ion{H}{2}~regions would have a maximum H$\alpha$ luminosity
of just under $10^{39}$~ergs~s$^{-1}$. This is consistent with an
apparent upper cutoff in \ion{H}{2}~region luminosity of
$\sim10^{39}$~ergs~s$^{-1}$ noted by Kennicutt, Edgar, \& Hodge (1989)
in galaxies of type Sb and earlier.

We now apply the Tully-Fisher relation between deprojected \ion{H}{1}
linewidth $W_{R}^{i}$ and absolute blue magnitude $M_{B}^{b,i}$ as given
for field galaxies by Pierce \& Tully (1992):
\begin{equation}
M_{B}^{b,i} = -7.48(\log W_{R}^{i}-2.50)-19.55\pm0.14 .
\label{eq:tf}
\end{equation}
\noindent
Using the corrections prescribed by Tully \& Fouqu\'{e} (1985) for
turbulent motion and profile shape, our $W_{20}=190\pm10$~km~s$^{-1}$
equates to a true velocity width $W_{R}=155\pm9$~km~s$^{-1}$. When
deprojected, this becomes $W_{R}^{i=20}=453\pm26$~km~s$^{-1}$ or
$W_{R}^{i=33}=285\pm 17$~km~s$^{-1}$, from which
equation~(\ref{eq:tf}) predicts $M_{B}^{b,i=20} =-20.71\pm0.33$ and
$M_{B}^{b,i=33}=-19.21\pm0.33$.  The apparent blue magnitude of
NGC~1433, corrected for internal (Tully \& Fouqu\'{e} 1985) and
Galactic extinction, is also a function of inclination: $B_{T}^
{b,i=20}=10.68\pm0.04$ and
$B_{T}^{b,i=33}=10.65\pm0.04$. Consequently, the distances derived for
the two inclination limits differ by a factor of 2; ${\rm
R}(i=20^{\circ})=19.1\pm3.5$~Mpc, while ${\rm
R}(i=33^{\circ})=9.4\pm1.7$~Mpc.  For the reasons discussed above, we
therefore favor a mean inclination for the disk of NGC~1433 closer to
$30^{\circ}$ than to $20^{\circ}$. As a consequence, the outer
pseudoring of NGC~1433 is, in fact, almost circular in the plane of
the galaxy.  While it may seem odd that the prototypical ringed barred
galaxy would have an intrinsic outer pseudoring axial ratio so much
greater than the average, we note that the models of Schwarz (1981) do
predict that the stronger bar cases will produce more nearly circular
outer rings.

One potential complication that may invalidate this argument is the
extent to which non-circular gas motions may disrupt the relationship
between luminosity and maximum rotational velocity in a spiral galaxy
(Jacoby et al. 1992). As a comparison of
Fig.~\ref{f:n1433ir} and Fig.~\ref{f:n1433rot} shows, the maximum
rotational velocity in NGC~1433 is reached well inside the bar,
and does not drop much below this maximum,
even in regions far from resonance. The maximum residuals between the
observed velocity field and a model velocity field with an inclination
held fixed at $33^{\circ}$ are still no more than 20~km~s$^{-1}$. For
a disk inclined at $20^{\circ}$ with non-circular motions to emulate a
$33^{\circ}$ inclination disk in pure circular motion, our kinematical
analysis would need to be biased towards inferring unusually high
velocities.

 From the crowding of gas and stars around the inner ring major axis,
it is apparent that particles spend much of their time there by
orbiting slower, and therefore must speed across the ring minor axis
to compensate. Since the ROCUR analysis excludes points within
$\pm25^{\circ}$ of the kinematic minor axis, which is almost
perpendicular to the apparent major axis of the inner ring, we could
indeed be favoring unduly large rotation velocities. However, the
situation is reversed in the outer pseudoring, and yet the rotational
velocity is observed to be only $\sim20$~km~s$^{-1}$ lower there than near
the inner ring. Thus, we conclude that non-circular motions, while not
an insignificant factor, do not corrupt the Tully-Fisher relation to the
extent of making the disk appear some $10^{\circ}$ closer to face-on
than it really is. Still, it is important to note that the Pierce \&
Tully calibration in equation~\ref{eq:tf} is based on galaxies having
inclinations $i\geq45^{\circ}$, and that applying it to galaxies of
much lower inclination involves, in a sense, an uncertain extrapolation.

Although the outer pseudoring of NGC~1433 may be more circular than
the average outer ring or pseudoring feature in barred galaxies, the
inner pseudoring of the galaxy is intrinsically {\it much more}
elongated than the average inner ring or pseudoring feature. For $i =
33^{\circ}$, the intrinsic axis ratio of the inner pseudoring of NGC~1433
is 0.61 (B86), compared to an average of $0.81\pm0.06$ obtained from
the analysis of CSRG data. Thus, NGC~1433 demonstrates how much an
individual object can depart from the averages obtained in the CSRG.

\subsection{\label{s:test}Testing the Resonance-Ring Theory}

\subsubsection{\label{s:n1433rrt}NGC~1433}

One of the primary motivations for this study is to use a knowledge of
the resonance orbit locations, coupled with the rotation curves, to
get a direct measure of the bar pattern speeds in NGC~1433 and
NGC~6300.  The bar pattern speed is clearly of great importance, since
theory predicts that it plays a key role in dictating not only the
positions of the resonances, but also how many and what types of
resonances will be present (Schwarz 1984; Byrd et~al. 1994, hereafter
BRSBC). Bar pattern speeds have been estimated in the past (B86;
van~Driel et al. 1988; van~Driel \& Buta 1991; Moore \& Gottesman
1995; JvM) for a few spiral galaxies, but nearly always, assumptions
have had to be made about the shape of the rotation curve, or the
location of corotation. As the summary in Table~\ref{t:bps} shows, bar
pattern speeds measured so far in such galaxies range between
approximately 20 and 50~km~s$^{-1}$.

With the information gathered in this study, and from previous photometric
analyses (B86; B87), we are now in a position to bypass such
assumptions, and use the theory of resonance ring formation to measure
the bar pattern speed $\Omega_{\rm p}$ directly. In the case of a weak
bar, the epicyclic frequency $\kappa$ is given by linear perturbation theory
as
\begin{equation}
\kappa = \left( R \frac{d\Omega^{2}}{dR} + 4\Omega^{2} \right)^{1/2}
\label{eq:bt}
\end{equation}
\noindent
(e.g., Binney \& Tremaine 1987). In the absence of non-circular
motions, $\Omega=V/R$, so that
\begin{equation}
\kappa = \sqrt{2}\frac{V}{R}\left(1+\frac{R}{V}\frac{dV}{dR}\right)^{1/2}.
\label{eq:kappa}
\end{equation}
\noindent
Using the rotation curve in Fig.~\ref{f:n1433rot}, we have plotted the
radial variation of the angular frequency $\Omega$, as well as the
Lindblad precession frequencies $\Omega+\kappa/2$, $\Omega-\kappa/2$,
and $\Omega-\kappa/4$ in Figure~\ref{f:n1433pf} (we have taken
$(dV/dR)$ to be the average of the slope to either side of each point
on the rotation curve). If the (linear) theory of resonance rings is
correct, then each of the resonance curves ought to intersect the
vertical line marking the radius of their associated ring at the same
frequency value, namely the bar pattern speed. Based on the mean ring
radii determined by B86, we derive $\Omega_{\rm p} =3\pm1$,
$22.4\pm0.4$, and $26.2\pm0.4$~km~s$^{-1}$~kpc$^{-1}$ from the nuclear
ring, the inner ring, and the outer ring, respectively.

The nuclear ring value is based on an inwards extrapolation of the
$\Omega-\kappa/2$ curve, and clearly, the effects of beam smoothing in
a region with such a steep velocity gradient make this estimate most
unreliable. In addition, the linear theory outlined above for the
radial variation of $\Omega$ and $\kappa$ may not be appropriate in
the region of a strong bar, and any associated gas inflow. Unlike the
outer pseudoring, which we have surmised to be almost circular, the
true axial ratio of the inner ring, being close to $\sim0.6$,
makes it difficult to assign a single characteristic radius for this
ring. Thus, we assign more weight to the $\Omega_{\rm p}$ implied by
the outer pseudoring, although there is still the likelihood that none
of the rings lies {\em exactly} at a resonance. The current theories,
however, stop short of predicting precisely where rings will form
relative to the resonant orbit locations.

Bearing all these uncertainties in mind (inclination, ring locations
relative to resonances, non-linear effects, etc.), we therefore
conservatively propose a value for the bar pattern speed (at the
current epoch) in NGC~1433 of $\Omega_{\rm p}=26\pm5$~km~s$^
{-1}$~kpc$^{-1}$.  This would place corotation at $R\sim6$~kpc, at
about 1.5~bar radii, and just beyond the inner ring.  From an orbit
analysis of the inner ring kinematics only, B86 came up with virtually
the same results by assuming a constant rotation velocity across the
ring region, although as Figs.~\ref{f:n1433ir} and \ref{f:n1433rot}
show, this is not strictly the case. We can also now confirm that the
gas-deficient regions seen in Fig.~\ref{f:n1433h1} between the inner
and outer \ion{H}{1} rings must be related to the vacant ``banana''
orbits near the $L_{4}$ and $L_{5}$ points at corotation, as predicted
by the strong-bar models of Schwarz (1981) and Contopoulos \&
Papayannopoulos (1980).

BRSBC used test-particle simulations to improve upon the earlier work
of Schwarz (1981, 1984) by testing the effects of bar pattern speeds,
bar strength, and rotation curve form on the number and location of
rings formed in the evolving models.  Using a combination of a flat
rotation curve (not unlike Fig.~\ref{f:n1433rot}), a ``strong'' bar
(relative radial force due to the bar of 20\%), and a ``medium to
slow'' bar domain pattern speed, they were able to produce after just
4 bar rotations a gas particle distribution with an uncanny
resemblance to our \ion{H}{1}~map of NGC~1433 (compare our
Fig.~\ref{f:n1433h1} to the lower right panel of their Fig.~12(b)),
even to the extent of reproducing gas peaks near the ``plume''
features seen optically. Although Table~\ref{t:bps} provides a much
too limited sample from which to draw any solid conclusions, it seems
not unreasonable that the ``slow to medium'' domain of bar pattern
speed would include our $\Omega_{\rm p}\sim 26$~km~s$^{-1}$ for
NGC~1433, with perhaps the ``fast'' pattern speed domain therefore
encompassing values of $\Omega_{\rm p}$ of 50--60~km~s$ ^{-1}$ and
upwards.

\subsubsection{\label{s:n6300rrt}NGC~6300}

Having found satisfactory agreement with the theory in the case of
NGC~1433, we next turn our attention to NGC~6300.
Figure~\ref{f:n6300pf} shows the precession frequency curves for
NGC~6300, based purely on the optical rotation curve data from
B87 for $R<9.3$~kpc, and on our \ion{H}{1} rotation curve
thereafter. Owing to the vastly better sampling of points in the inner
disk of NGC~6300 {\em cf.} NGC~1433, the radius axis has been plotted
on a logarithmic scale. The vertical dashed lines here mark not
separate rings, but the same approximate boundaries for the broad
inner ring as given by B87, and marked in
Fig.~\ref{f:n6300ir}. By again associating this inner ring feature
with the inner second harmonic resonance, we would infer a
bar pattern speed for NGC~6300 in the broad range $\Omega_{\rm p}=27
\pm 8$~km~s$^{-1}$. Thanks to the extended radial coverage from our
\ion{H}{1}~data, we would therefore expect to see some sign of an
outer ring or pseudoring somewhere between $R=8-14$~kpc, where the
$\Omega_{\rm p}$ limits intersect the $\Omega+\kappa/2$ curve.
Figure~\ref{f:n6300ir} reveals no such feature in this range ($115'' <
r < 200''$), the bump at $r=145''$ already having been ascribed to
short spiral segments, and not a complete ring (\S~\ref{s:n6300h1d}).

In the same vein, these $\Omega_{\rm p}$ limits based on the inner
ring also intersect the $\Omega-\kappa/2$ locus at a number of radii.
According to the resonance-ring theory, there ought therefore to be at
least one nuclear ring, probably inside $R=0.8$~kpc. Such a ring would
be no bigger than our synthesized beamwidth, and would not be apparent
in Fig.~\ref{f:n6300ir}. No optical nuclear ring was reported by
B87, although such a ring could conceivably be camouflaged
by the foreshortened bar, dust lanes, or the type~2 Seyfert nature
of the nucleus (Phillips, Charles, \& Baldwin 1983).

We are therefore left to conclude that either the resonance-ring
theory for the origin of stellar and gaseous rings (which performed
creditably in the case of NGC~1433) is flawed, or else the ring in
NGC~6300 may not be a feature directly driven by the bar. There is
reason to believe that the latter situation is a definite possibility.
B87 noted that when NGC~6300 is approximately deprojected, the inner
pseudoring is aligned neither parallel nor perpendicular to the bar,
but at an intermediate angle. Sygnet et~al. (1988) have
suggested that NGC~6300 is a multiple pattern speed galaxy because of
this misalignment. Sellwood \& Sparke (1988) have suggested
that spiral patterns in barred galaxies may not, in general, be driven
by the bar, but are independent patterns with a slower pattern speed
than the bar. Tagger et~al. (1987), Sygnet et~al.
(1988), and Sellwood \& Sparke (1988) have discussed the
idea of mode-coupling between the spiral and bar patterns. In this
scenario, the corotation radius of the bar coincides with the radius
of the inner Lindblad resonance (ILR) of the spiral. As a consequence
of the different pattern speeds, one would expect a random
distribution of phase difference between the ends of the bar and the
beginning of the spiral arms. In the case of NGC~6300, we note that
the spiral structure does not begin at the ends of the bar, but about
$30^{\circ}$ ahead of the bar. However, Sellwood \& Sparke
(1988) present simulations showing that even with different pattern
speeds, the spiral and the bar will often appear smoothly connected.

It is possible for an inner resonance-ring to be composed of distinct
segments, like that of NGC~1433, but the \ion{H}{1} ``spurs''
extending from the gas ring of NGC~6300 indicate that the spiral arm
segments in NGC~6300 are much less tightly confined than those in
NGC~1433. Still another difference between NGC~6300 and NGC~1433 is
that the peaks in the \ion{H}{1}~ring of NGC~6300 do lie along the ring
major axis (partly due to projection effects), but this then places
them almost perpendicular to the bar axis, rather than along the bar
axis as is the case in NGC~1433. Finally, despite apparently having a
fast bar powerful enough to drive gas inflows, there are no relatively
empty regions near corotation in NGC~6300 analogous to those in
NGC~1433.

There is a third possibility, originally outlined by B87,
which is that the tightly-wound spiral arms in NGC~6300 are still in
the process of forming a closed, continuous resonance ring. It takes
time for the torques exerted by the bar to stretch spiral features
into rings, and continual gas infall or recycling can retard this
evolution. The presence of an extended, warped \ion{H}{1}~disk and
gas tail in NGC~6300 may point to ongoing gas accretion in this
galaxy. The simulations of BRSBC have shown how
two galaxies with nearly the same pattern speed (such as NGC~1433 and
NGC~6300) can still appear morphologically quite distinct, if only
because one galaxy was slower to develop a bar asymmetry (not to
mention the possibility that bars may be destroyed and reformed many
times during a Hubble time; Sellwood \& Wilkinson 1993).

Further support for the evolution scenario is provided by
Figure~\ref{f:h1onha}, which shows contours of the uniform-weighted
\ion{H}{1} surface density on the H$\alpha$ image used in \S~\ref{s:n6300c}.
When compared with Fig.~\ref{f:n6300h1onb}, it is clear that the recent
massive star formation in NGC~6300 follows the \ion{H}{1}~ring much
better than does the older and slightly smaller stellar pseudoring.
Thus, the stellar orbits (which are slower to react to the bar torques
than the gaseous component) could be stretched still further in the next
few Gyr, although any further slowdown in the bar pattern speed will
cause the stellar ring to take even longer to reach its final
mature stage. This evolution may also be associated with a
transition from a disk-dominated to a (nucleus $+$ bar)-dominated
radio continuum source.

Finally, an outer ring will take much longer to form than an inner
ring, while the nuclear ring will form quickest of all
($\sim10^{7}-10^{8}$~years; Combes 1993). Thus, if a
nuclear ring could be confirmed in NGC~6300, perhaps from near-IR
imaging, then the case for interpreting NGC~6300 as a barred spiral
evolving into a resonance-ring galaxy (with NGC~1433 representing the
pinnacle of this evolution) would be further strengthened.

The nearby barred spiral galaxy NGC~1365 possesses an outer pseudoring
(CSRG), and shares some of the \ion{H}{1}~characteristics of both
NGC~1433 and NGC~6300. In a recent comprehensive VLA \ion{H}{1} study,
JvM found that just like in NGC~1433, the \ion{H}{1} disk in NGC~1365
does not extend much beyond the optical disk, and there is an almost
complete \ion{H}{1}~ring visible in their low-resolution map at the
radial distance of the optical pseudoring. In other ways however,
NGC~1365 has more in common with NGC~6300, such as the presence of a
warp and an asymmetric extension to the gas disk. By associating
dust-lane arm crossings with corotation, and minima in the streaming
motions along the arms with individual resonances, JvM used a not
dissimilar analysis to ours to arrive at a bar pattern speed of
$\Omega_{\rm p}=21\pm1$~km~s$^{-1}$. However, this would place the
outer Lindblad resonance at $R\sim208''$, barely \twothirds\ of the radial
distance of the optical and \ion{H}{1}~pseudoring. In addition,
despite the obvious presence of a strong bar, the $L_{4}$ and $L_{5}$
regions have not been evacuated to the same degree as for NGC~1433.
Thus, unless NGC~1365 possesses multiple pattern speeds, its status as
a {\em bona fide} resonance-ring galaxy akin to NGC~1433 is still open
to question.

\section{Conclusions}

In this paper we have presented the distribution and velocity fields of
neutral hydrogen in NGC~1433 and NGC~6300, two of the closest ringed
barred spirals in the southern sky. Though classified similarly
with regard to morphology, the two objects are quite dissimilar in
many respects. We summarize our main results for each object separately.
For NGC 1433 we have found:
\begin{enumerate}
\item An \ion{H}{1}~distribution that follows the optical structure very
      closely. The inner and outer pseudorings are very well-defined in the
      \ion{H}{1}~map, as well as the two secondary arcs (or ``plumes'') off
      the leading sides of the inner ring. There is also central emission
      with a peak slightly to the east of the nucleus. The bar region, and
      the areas between the inner and outer rings along the bar minor axis
      line, are deficient in neutral atomic hydrogen.
\item Each ring feature in NGC~1433 is associated with a maximum in the
      azimuthally-averaged gas surface density. The plumes are only local
      maxima at their radii.
\item If the outer isophotes of NGC 1433~represent an axisymmetric region,
      then the inclination of the system must be close to $33^{\circ}$.
      However, if the outer pseudoring is intrinsically oval, as is
      typically the case, then the inclination could be 20$^{\circ}$ or
      less. Application of the Tully-Fisher relation to the \ion{H}{1}~data
      favors the higher estimate of the inclination, meaning that in spite
      of the strong bar, the outer pseudoring of NGC~1433 is nearly circular
      in the disk plane. On the other hand, this also implies that the inner
      ring of NGC~1433 is intrinsically much more elongated than the average.
\item The strength of the (non-thermal) radio continuum emission in NGC~1433
      is well correlated with the local density of \ion{H}{2}~regions, with
      roughly equal contributions coming from each end of the bar and from
      the nucleus. Although a definite detection is far from certain, we
      have been able to place an upper limit on the present 1.38~GHz flux
      density from the Type~II SN~1985P as $S_{\rm 1.38~GHz} < 0.5$~mJy.
\item By identifying the outer pseudoring with the outer Lindblad
      resonance, the \ion{H}{1}~rotation curve of NGC~1433 gives a bar
      pattern speed of $26\pm5$~km~s$^{-1}$~kpc$^{-1}$. This pattern speed is
      consistent with the location of the inner 4:1 resonance being near the
      position of the inner ring, but our observations lack the spatial
      resolution to test for an association of the nuclear ring with the
      inner Lindblad resonance. The gas deficient regions between the two
      rings appears to be associated with the Lagrangian points $L_4$ and
      $L_5$, which are located at corotation along the minor axis line of
      the bar. The orbits in these regions are unstable in the presence of
      a strong bar. Thus, the structure of NGC~1433 compares favorably with
      test particle models of barred spirals.
\end{enumerate}
For NGC~6300 we have found the following:
\begin{enumerate}
\item The \ion{H}{1}~distribution extends well beyond the optical disk.
      This is very unlike NGC~1433, and highlights a fundamental difference
      between the two systems. Like NGC~1433, however, the inner pseudoring
      is a region of gas concentration. There are no deficiencies of gas
      in the regions outside the inner ring where the Lagrangian points
      $L_4$ and $L_5$ might be expected to be found.
\item Throughout most of the optical disk, the azimuthally averaged surface
      density of \ion{H}{1}~gas in NGC~6300 is higher than in NGC~1433.
      The maximum surface density is found near the location of the inner
      pseudoring, and the density peak is broader than the extent of the
      optical ring.
\item The extended \ion{H}{1}~disk of NGC~6300 exhibits both a loose ``tail''
      feature on its western edge, and a $20^{\circ}$ warp in its kinematic
      line-of-nodes. The overall \ion{H}{1}~mass-to-blue luminosity ratio is
      not, however, unlike that for NGC~1433 or for Sab galaxies in general.
\item We confirm the existence of non-circular gas motions in the bar region,
      as hinted at by optical emission-line spectroscopy, but such motions
      seem to be confined within the inner \ion{H}{1}~ring. There appears to
      be a discrepancy between the optical rotational velocity curve derived
      from a series of spectra at various position angles, and the rotation
      curve derived from the global \ion{H}{1}~velocity field, which is most
      apparent in the region of the inner pseudoring.
\item Emission from within the inner ring of NGC~6300 totally dominates even
      the contribution from the type~2 Seyfert nucleus to the 1.38~GHz
      continuum.
\item On the assumption that the inner pseudoring in NGC~6300 is associated
      with the inner 4:1 resonance, we find $\Omega_{\rm p}=27\pm8$~km~s$^
      {-1}$~kpc$^{-1}$. However, there is no trace optically or in the
      \ion{H}{1} of either an outer pseudoring or a nuclear ring at the
      positions implied if this is indeed the case. While it may be that
      the inner pseudoring of NGC~6300 has little to do with resonance
      phenomena at all, we have also presented several arguments for why
      we believe NGC~6300 may represent merely a ``juvenile'' stage in the
      transition from a barred spiral to a fully-fledged resonance-ring
      galaxy such as NGC~1433.
\end{enumerate}

Although this study has provided one of the first opportunities to directly
compare real examples of resonance-ring galaxies with numerical models,
there is still much work that needs to be done both observationally and
theoretically. As Table~\ref{t:bps} attests, the sample of galaxies for which
direct measurements of bar pattern speeds are available is still somewhat
limited for examining the influence of pattern speed domain on galactic
structure. Although the linear perturbation theory used in this analysis
may not be strictly appropriate in the central regions of these galaxies
in the presence of non-circular motions, the key limiting factors are still
a lack of spatial resolution, and no specific predictions of where the gas
and stellar rings ought to lie relative to the resonant orbit locations.
Even so, many key predictions of the Schwarz (1981) and
BRSBC models have been borne out by these new observations,
and we may even have the beginnings of an evolutionary sequence for
resonance-ring galaxies.

% If nothing else however, the striking \ion{H}{1}~morphology of NGC~1433
% revealed by the ATCA in this study must surely earn it the undisputed title
% of ``Lord of the (Resonance) Rings.''

\acknowledgments

We are grateful to D.~Crocker and N.~Killeen for installing and
maintaining the {\sc aips} software at the University of Alabama, and
to V.~McIntyre for producing Figure~\ref{f:beamplot}. We wish to thank
G.~Purcell, D.~Crocker, and M.~Mazzolini for supplying data in advance
of publication, and acknowledge useful comments from G.~Byrd,
B.~Koribalski, and an anonymous referee. This research has made use of
the NASA/IPAC Extragalactic Database (NED), which is operated by the
Jet Propulsion Laboratory, Caltech, under contract with the National
Aeronautics and Space Administration. H.~T.~acknowledges support from
CONACYT through a graduate student scholarship and DGAPA/UNAM through
grant IN107094. This work has been supported by EPSCoR grant
EHR-9108761.

% That's the end of the main body of the paper.  Now we will have some
% back matter.

%\clearpage

\begin{deluxetable}{lcc}
\tablewidth{0pt}
\tablecaption{\label{t:obspar}Observational and Derived Parameters
for NGC~1433 and NGC~6300}
\tablehead{
\colhead{Galaxy} & \colhead{NGC~1433} & \colhead{NGC~6300}
}
\startdata
Hubble type\tablenotemark{a}      & (R$^{\prime}_{1}$)SB(\underline{r}s)ab &
                                    SB(\underline{r}s)b \nl
Configurations                    & 0.375, 0.75A, 0.75B, 1.5C &
                                    0.375, 0.75A, 0.75B, 1.5C, 1.5D \nl
Total observation time            & 48~hrs           &     58~hrs \nl
$\alpha$ (J2000 pointing center)  & $03^{\rm h}\, 42^{\rm m}\, 01\fs2$ &
                                    $17^{\rm h}\, 16^{\rm m}\, 58\fs2$ \nl
$\delta$ (J2000 pointing center)  & $-47^{\circ}\, 13'\, 18''$ &
                                    $-62^{\circ}\, 49'\, 11''$ \nl
FWHP synthesized beam\tablenotemark{b} &  $33''\times 27''$ ($1.86\times 1.52$
                           kpc) &  $25''\times23''$ ($1.72\times1.58$~kpc) \nl
Channel map r.m.s.~noise\tablenotemark{b} &  0.9~mJy beam$^{-1}$ &
                                   1.0~mJy~beam$^{-1}$ \nl
Continuum map r.m.s.~noise\tablenotemark{b} & 0.14~mJy beam$^{-1}$ &
                                   0.15~mJy~beam$^{-1}$ \nl
Distance adopted\tablenotemark{c}  & 11.6~Mpc & 14.3~Mpc \nl
Inclination adopted                & $33^{\circ}$ & $(51\pm4)^{\circ}$ \nl
$M_{B}^{b,i}$                      & $-19.21\pm0.33$ & $-19.92\pm0.24$ \nl
$M_{\rm H\,I}$                     &  $(1.05\pm0.05)\times10^{9}$~M$_{\odot}$
                                    & $(2.6\pm0.2) \times 10^{9}$~M$_{\odot}$
                                     \nl
$M_{\rm H\,I}/L_{B}$               & $0.14\pm0.05$~M$_{\odot}/L_{\odot}$ &
                                    $0.18\pm 0.06$~M$_{\odot}/L_{\odot}$ \nl
\enddata
\tablenotetext{a}{From the {\em Catalog of Southern Ringed Galaxies},
                  Buta 1995}
\tablenotetext{b}{Uniform weighting}
\tablenotetext{c}{Tully 1988}
\end{deluxetable}

\begin{deluxetable}{lccc}
%\tablewidth{330pt}
\tablecaption{\label{t:profpars}Global \protect{\ion{H}{1}} Profile Parameters}
\tablehead{
\colhead{Profile source} & \colhead{Integral Flux Density} &
\colhead{Systemic Velocity} & \colhead{$W_{20}$} \\
\multicolumn{1}{c}{~~} & \multicolumn{1}{c}{(Jy~km~s$^{-1}$)} &
\multicolumn{1}{c}{(km~s$^{-1}$)} & \multicolumn{1}{c}{(km~s$^{-1}$)}
}
\tablecolumns{4}
\startdata
\cutinhead{NGC~1433}
Natural  & 34.4 & 1075 & 191 \nl
Uniform  & 32.2 & 1075 & 188 \nl
Parkes (This study) & 31.4 & 1074 & 191 \nl
Parkes (Reif et al. 1982) & 23.6 & 1080 & 191 \nl
\cutinhead{NGC~6300}
Natural  & 55.0 & 1110 & 317 \nl
Uniform  & 51.0 & 1111 & 314 \nl
Parkes (Mazzolini et al. 1995) & 57.2 & 1110 & 320 \nl
Parkes (Reif et al. 1982) & 61.0 & 1110 & 341 \nl
\enddata
\end{deluxetable}

\begin{deluxetable}{lcl}
\tablewidth{330pt}
\tablecaption{\label{t:bps}Bar Pattern Speeds}
\tablehead{
\colhead{Galaxy} & \colhead{$\vert\Omega_{\rm p}\vert$ (km~s$^{-1}$~kpc$
^{-1}$)} & \colhead{Reference}
}
\startdata
% NGC~936 & $60\pm14$ & Kuijken 1995 \nl
NGC~1365 & $21\pm1$ & J\"{o}rs\"{a}ter \& van~Moorsel 1995 \nl % No ring
NGC~1398 & $48\pm15$\tablenotemark{a} & Moore \& Gottesman 1995 \nl % IR
NGC~1433 & $28\pm7$\tablenotemark{b} & Buta 1986 \nl % NR:IR:OP
NGC~1433 & $26\pm5$\tablenotemark{b} & This study \nl % NR:IR:OP
NGC~5101 & 19~~~~~~  & van~Driel et al. 1988 \nl     % NL:IR:OP
NGC~6217 & $41\pm3$  & van~Driel \& Buta 1991 \nl    % IP:OP
NGC~6300 & $27\pm8$\tablenotemark{c} & This study \nl % IP
\enddata
% NGC~7217 & 80     & Bosma 1995 \nl
\tablenotetext{a}{Based on a distance of 16.1~Mpc (Tully 1988), and the
uncertainty in associating the ring with a particular resonance.}
\tablenotetext{b}{Based on a distance of 11.6~Mpc (Tully 1988).}
\tablenotetext{c}{Assuming the inner ring in NGC~6300 corresponds to the
$\Omega-\kappa/4$ resonance.}
\end{deluxetable}

%\clearpage
% Now comes the reference list.  In this document, we used \markcite to call
% out citations, so we must use \reference in the reference list.

%%

% N.B.!!!!! \ion{}{} is fragile, and therefore needs to be enclosed in a
%\protect{}

\clearpage

\begin{figure}
%\plotone{beamplot.ps}
\figcaption{\label{f:beamplot}({\em left}) Major-axis radial profiles of the
dirty (natural-weighted) beam and the best-fitting Gaussian clean beam used
by {\sc aips}, and ({\em right}), the integrated flux within a given radius
from each beam shape. The implications of the differences between the dirty
and clean beams for image cleaning and total flux estimates are discussed
in the paper by J\"{o}rs\"{a}ter \& van~Moorsel (1995).}
\end{figure}

\begin{figure}
%\plotfiddle{ngc1433.ps}{14cm}{0}{80}{80}{-250}{-120}
\figcaption{\label{f:n1433cm}\protect{\ion{H}{1}} channel maps of NGC~1433
after CLEANing with a uniform-weighted beam. Heliocentric velocities in
the optical convention are given in km~s$^{-1}$. The darkest grey level
corresponds to a flux density $\sim15$~mJy~beam$^{-1}$.}
\end{figure}

\begin{figure}
%\plotfiddle{naunpks1433.ps}{14cm}{270}{100}{100}{-300}{500}
\figcaption{\label{f:n1433prof}Global \protect{\ion{H}{1}} profiles of
NGC~1433 obtained by measuring the pure emission within each plane of
the uniform (dashed line) and natural (solid line) weighted cubes,
corrected for primary beam attenuation. For comparison, the dash-dot
line is a recent spectrum from a single pointing with the Parkes
64~m antenna, not corrected for beam dilution.}
\end{figure}

\begin{figure}
%\plotone{N1433H1C.PS}
\figcaption{\label{f:n1433h1}Grey-scale and contours of \protect{\ion{H}{1}}
column density in NGC~1433. Contours represent (projected) column densities
of 1.4, 1.9, 2.5, 3.2, 3.8, 5.0, and $6.3\times10^{20}$~cm$^{-2}$. The FWHP
of the synthesized beam is shown by the ellipse in the bottom left corner.
The cross marks the position of the dynamical center.}
\end{figure}

\begin{figure}
%\plotone{H1ONB.PS}
\figcaption{\label{f:h1onb}Contours of the \protect{\ion{H}{1}} column density
overlaid on a $B$-band image of NGC~1433. The contour levels are the same
as for Fig.~\protect{\ref{f:n1433h1}}, but have been regridded to match the
pixel scale of the optical image. The $B$-band intensity has been displayed
on a logarithmic scale, in order to make both the outer pseudoring and
nuclear detail visible.}
\end{figure}

\begin{figure}
%\plotfiddle{n1433ir.ps}{10cm}{0}{90}{90}{-300}{-100}
\figcaption{\label{f:n1433ir}Radial distribution of the azimuthally-averaged
\protect{\ion{H}{1}} surface density, corrected to face-on, assuming a
disk inclination of $33^{\circ}$ and optically thin gas. The mean radii of
prominent optical features, as measured by Buta (1986b) are
indicated by the dashed lines. The bar at top right denotes the approximate
FWHP beamwidth.}
\end{figure}

\begin{figure}
%\plotfiddle{N1433VF.PS}{14cm}{0}{80}{80}{-250}{-120}
\figcaption{\label{f:n1433vf}Isovelocity contours overlaid on the
\protect{\ion{H}{1}} column density map of NGC~1433. The contours are
at intervals of 10~km~s$^{-1}$ and major velocity contours are marked.
The closed contours along the kinematic major axis correspond to heliocentric
velocities of 990~km~s$^{-1}$ ({\em top}) and 1160~km~s$^{-1}$ ({\em bottom}).
Many spurious velocity points around the edge of the disk and around the vacant
southern region have been blanked out, due to poor signal-to-noise.
The cross marks the position of the dynamical center, as determined from the
ROCUR analysis (see text).}
\end{figure}

\begin{figure}
%\plotfiddle{n1433rot.ps}{10cm}{0}{90}{90}{-300}{-100}
\figcaption{\label{f:n1433rot}\protect{\ion{H}{1}} rotational velocities
derived from the velocity field, assuming circular motion in rings
inclined at $33^{\circ}$, with rotation center, systemic velocity, and
position angle held fixed at the values given in the text. The points
are plotted at 1~beamwidth intervals, based on separate fits to the
receding (i.e., mostly the southern) half, the approaching (northern)
half, and to the entire disk.  The formal fitting errors are usually
much less than the differences between these three fits.}
\end{figure}

\begin{figure}
%\plotone{N1433CONHA.PS}
\figcaption{\label{f:n1433conha}Contours of
1.38~GHz continuum flux density overlaid on a 1200~s
continuum-subtracted H$\alpha$ image obtained by D.~Crocker and
G.~Purcell with the CTIO 1.5~m telescope. The radio continuum contours
are at levels of 0.4, 0.6, 0.8, 1.0, 2.0, and 3.0~mJy beam$^{-1}$. The
cross $1'$ northeast of the nucleus marks the approximate optical
position of the Type~II supernova SN~1985P discovered by R.~Evans.}
\end{figure}

\begin{figure}
%\plotfiddle{ngc6300.ps}{12cm}{270}{70}{70}{-330}{450}
\figcaption{\label{f:n6300cm}\protect{\ion{H}{1}} channel maps of NGC~6300
after CLEANing with a uniform-weighted beam. Heliocentric velocities in
the optical convention are given in km~s$^{-1}$. The darkest grey level
corresponds to a flux density $\sim40$~mJy~beam$^{-1}$.}
\end{figure}

\begin{figure}
%\plotfiddle{naunpks6300.ps}{14cm}{270}{100}{100}{-300}{500}
\figcaption{\label{f:n6300prof}Global \protect{\ion{H}{1}} profiles of NGC~6300
obtained by measuring the pure emission within each plane of the
uniform (dashed line) and natural (solid line) weighted cubes,
corrected for primary beam attenuation. For comparison, the dash-dot
line is a spectrum from a single pointing with the Parkes 64~m
antenna, which has not been corrected for beam dilution.}
\end{figure}

\begin{figure}
%\plotfiddle{N6300NAH1.PS}{14cm}{270}{80}{80}{-330}{450}
\figcaption{\label{f:n6300h1}Grey-scale and contours of \protect{\ion{H}{1}}
column density in NGC~6300, with natural weighting applied to the
{\em uv}-data. Contours represent (projected) column densities
of 0.1, 0.5, 1.4, 3.7, 5.5, 7.3, and $9.2\times10^{20}$~cm$^{-2}$. The FWHP
of the synthesized beam is shown by the ellipse in the bottom left corner.
The cross marks the position of the dynamical center.}
\end{figure}

\begin{figure}
%\plotone{N6300H1ONB.PS}
\figcaption{\label{f:n6300h1onb}Contours of the \protect{\ion{H}{1}}
column density overlaid on a $B$-band image of NGC~6300. The contour
levels are at projected column densities of 0.6, 2.0, 4.9, 7.9, and
$10.8 \times 10^{20}$~cm$^{-2}$. The optical nucleus appears to be
elongated, due to the presence of a bright foreground star due west
of the nucleus, and another just to the southeast.}
\end{figure}

\begin{figure}
%\plotfiddle{n6300ir.ps}{10cm}{0}{90}{90}{-300}{-100}
\figcaption{\label{f:n6300ir}Radial distribution of the azimuthally-averaged
\protect{\ion{H}{1}} surface density, corrected to face-on, for which the
shape of each ring (inclination and position angle) is set by the rotation
curve fitting (Fig.~\protect{\ref{f:n6300rot}}). The approximate inner and
outer boundaries of the inner ring, as given by Buta (1987), are
indicated by the dashed lines. The bar at top right denotes the approximate
FWHP beamwidth.}
\end{figure}

\begin{figure}
%\plotfiddle{N6300NAVF.PS}{14cm}{270}{80}{80}{-330}{450}
\figcaption{\label{f:n6300vf}Isovelocity contours overlaid on the
\protect{\ion{H}{1}} column density map of NGC~6300, both formed from a
moment analysis of the natural-weighted {\em uv}-dataset. The contour
interval is 20~km~s$^{-1}$, and some major velocity contours are marked.
The cross marks the position of the dynamical center.}
\end{figure}

\begin{figure}
%\plotfiddle{N6300BARVF.PS}{14cm}{270}{75}{75}{-300}{450}
\figcaption{\label{f:n6300bar}Close-up of isovelocity contours in the
bar and inner ring region of NGC~6300 overlaid on an \protect{\ion{H}{1}}
column density map, both formed from a moment analysis of the
uniform-weighted {\em uv}-dataset. The contour interval is 20~km~s$^{-1}$,
and the cross marks the position of the dynamical center.}
\end{figure}

\begin{figure}
%\plotfiddle{n6300rot.ps}{14cm}{0}{60}{60}{-200}{-60}
\figcaption{\label{f:n6300rot}Kinematical analysis results for the
uniform-weighted velocity field, assuming circular motion in rings
inclined at angle $i$, having line-of-nodes position angle $\theta$,
and with rotation center and systemic velocity fixed. The points are
plotted at 1~beamwidth intervals, based on separate fits to the receding
(i.e., mostly the eastern) half, the approaching (western) half, and to
the entire disk. The open circles in the rotational velocity plot are the
optical ionized gas rotational velocities computed by Buta (1987), with
fixed $\theta=108^{\circ}$ and $i=52^{\circ}$.}
\end{figure}

\clearpage
\begin{figure}
%\plotone{N6300CONHA.PS}
\figcaption{\label{f:n6300conha}Contours of
1.38~GHz continuum flux density overlaid on a 600~s
continuum-subtracted H$\alpha$ image obtained by D.~Crocker and
G.~Purcell with the CTIO 1.5~m telescope. The radio continuum contours
are at levels of 1.0, 1.3, 1.5, 1.7, 1.9, 2.2, 2.4, 2.6, 3.0, 3.4,
4.5, 5.4, 6.0, 8.0, 11.0, 12.5 and 15.0~mJy~beam$^{-1}$. The two
extensions to the south are artifacts, and should be ignored. The four
distinct objects around the edge of this image have flux densities
(going clockwise from top right) of 28, 3.3, 2.5, and 4.7~mJy. }
\end{figure}

\begin{figure}
%\plotfiddle{n1433pf.ps}{10cm}{0}{90}{90}{-300}{-100}
\figcaption{\label{f:n1433pf}The various precession frequency curves for
NGC~1433 are shown as solid lines, with their respective frequencies marked.
The vertical dashed lines indicate the mean radii of the bar, plumes, and
rings, as in Fig.~\protect{\ref{f:n1433ir}}, and the horizontal dotted lines
mark the bar pattern speed implied if the rings do correspond to specific
resonances.}
\end{figure}

%\clearpage
\begin{figure}
%\plotfiddle{n6300pf.ps}{10cm}{0}{90}{90}{-300}{-100}
\figcaption{\label{f:n6300pf}The various precession frequency curves for
NGC~6300 are shown as solid lines, with their respective frequencies marked.
The vertical dashed lines indicate the approximate extent of the inner
ring, and the horizontal dotted lines mark the possible bar pattern speed
range implied if this ring was in fact associated with the 2HR$^{-}$
resonance. Note that the radius scale is plotted logarithmically to
accommodate the different sampling in the optical and \protect{\ion{H}{1}}
regimes.}
\end{figure}

\begin{figure}
%\plotone{H1ONHA.PS}
\figcaption{\label{f:h1onha}Contours of \protect{\ion{H}{1}} column density as
shown in Fig.~\protect{\ref{f:n6300h1onb}} overlaid on the same H$\alpha$
image as in Fig.~\protect{\ref{f:n6300conha}}. Notice the extremely good
agreement between both the gas morphology and column density when compared
with the distribution and brightnesses of the \protect{\ion{H}{2}}~regions,
and with the stellar ring (Fig.~\protect{\ref{f:n6300h1onb}}).}
\end{figure}

% That's all, folks.
%

\end{document}